\documentclass[12pt, preprint]{emulateapj}

\usepackage{natbib}
\usepackage{amsmath}

\shortauthors{Shi et al.}

\begin{document}

\title{A  Joint Model Of  X-ray  And Infrared  Backgrounds. II. Compton-Thick AGN Abundance
\footnotemark[*]}
\footnotetext[*]{Online calculators of the model are available at http://5muses.ipac.caltech.edu/5muses/EBL\_model/index.html}

\author{Yong  Shi\altaffilmark{1,2,3},  George Helou\altaffilmark{1},  Lee Armus\altaffilmark{1}} 

\altaffiltext{1}{Infrared  Processing and Analysis   Center,   California    Institute   of   Technology,   1200 E. California Blvd, Pasadena, CA 91125, USA}
\altaffiltext{2}{School of Astronomy and Space Science, Nanjing University, Nanjing 210093, China}
\altaffiltext{3}{Key Laboratory of Modern Astronomy and Astrophysics (Nanjing University), Ministry of Education, Nanjing 210093, China}

\begin{abstract}  

We estimate the abundance of Compton-thick (CT) active galactic nuclei (AGN) based on our joint
model of X-ray  and infrared backgrounds.  At $L_{\rm  rest 2-10 keV}$
$>$ 10$^{42}$  erg/s, the CT AGN  density predicted by our  model is a
few $\times$10$^{-4}$ Mpc$^{-3}$  from $z$=0 up to $z$=3.  CT AGN with
higher luminosity  cuts ($>$ 10$^{43}$, 10$^{44}$  \& 10$^{45}$ erg/s)
peak at  higher $z$ and  show a rapid  increase in the  number density
from $z$=0 to $z$$\sim$2-3.   The CT  to all AGN ratio appears to be
low  (2-5\%) at  $f_{\rm 2-10keV}$  $>$ 10$^{-15}$  erg/s/cm$^{2}$ but
rises  rapidly toward  fainter flux  levels.  The  CT AGN  account for
$\sim$ 38\% of the total accreted SMBH mass  and contribute $\sim$ 25\%
of the cosmic X-ray background spectrum at 20 keV.  Our model predicts
that the majority  (90\%) of luminous and bright  CT AGN ($L_{\rm rest
2-10  keV}$ $>$ 10$^{44}$  erg/s or  $f_{\rm 2-10keV}$  $>$ 10$^{-15}$
erg/s/cm$^{2}$) have detectable hot dust 5-10 $\mu$m emission which we
associate with a dusty torus.  The fraction drops for fainter objects,
to  around 30\%  at $L_{\rm  rest 2-10  keV}$ $>$  10$^{42}$  erg/s or
$f_{\rm 2-10keV}$  $>$ 10$^{-17}$ erg/s/cm$^{2}$.  Our model confirms
that heavily-obscured  AGN ($N_{\rm HI}$ $>$  10$^{23}$ cm$^{-2}$) can
be separated  from unobscured  and mildly-obscured ones  ($N_{\rm HI}$
$<$ 10$^{23}$ cm$^{-2}$) in the plane of observed-frame X-ray hardness
vs. mid-IR/X-ray ratio.

\end{abstract}


\section{Introduction}

Active   galactic  nuclei  (AGN)   with  Compton-thick   (CT)  nuclear
obscuration  ($N_{\rm HI}$ $>$  1.5$\times$10$^{24}$ cm$^{-2}$)  are a
crucial piece in the quest  for a complete census  of the AGN
population.  While {\it Chandra}  and {\it XMM-Newton} have revealed a
large   population  of   AGN   up  to   $z$$\sim$5  and   demonstrated
unambiguously   the  dominance   of  supermassive   black-hole  (SMBH)
accretion   in  the   obscured   phase  \citep{Mainieri02,   Perola04,
Hasinger05,  Barger05},  the necessity  of  having  a  high SNR  X-ray
spectrum including detection above and below rest-frame energies of 10
keV, largely limits the detectability to reveal the presence of CT AGN
in  deep surveys  \citep{Tozzi06, Georgantopoulos09}.   However, there
are several compelling  reasons to suspect an abundant  distant CT AGN
population: (1) in the local universe the CT AGN comprise roughly 50\%
of the optically-selected  AGN sample \citep{Risaliti99, Guainazzi05};
given  the dusty high-$z$  universe, the  distant CT  AGN may  be more
abundant.  (2) Cosmic X-ray  background (CXB) population models invoke
luminosity functions of AGN with different HI columns to fit the X-ray
survey data \citep{Comastri95, Gilli07, Treister09a}, which requires a
large number  of CT  AGN to  re-produce the CXB  spectrum at  its peak
(20-30  keV)  ; this  general  conclusion  is  largely independent  of
detailed   assumptions  in  the   model.   (3)   The  multi-wavelength
techniques  that combine  the  X-ray data  with optical/IR  photometry
offer  powerful  ways  to  identify  CT candidates,  and  indicate  an
increasing spatial density of CT AGN with redshift \citep{Alexander08,
Daddi07, Fiore09, Luo11, Treister09b}.

In \citet{Shi13} (hereafter Paper  I), we presented a joint population
model of X-ray  and infrared backgrounds that fits  the survey data in
the 0.5-60 keV and 24-1200 $\mu$m bands, with the goal of studying the
cosmic evolution of AGN and  dusty starbursts.  We discuss here the CT
AGN abundance derived from this  model.  In contrast to CXB models
that only  fit X-ray data primarily at energies  below 10 keV, our approach  is  fundamentally
different.   CXB  models usually use 0.1-10 keV data to fit the Compton-thin AGN
counts,  extrapolate the result  to  20-30 keV and then subtract this from 
the CXB spectrum to derive the abundance of CT AGN. Our   model    constrains  the
Compton-thin  AGN from  0.5-10 keV  data and  starburst  galaxies from
far-IR data,  and compares the 24  $\mu$m from these populations to the IR background and
known distributions of 24 $\mu$m-detected sources.  The
residual of the 24$\mu$m  emission after subtracting the contributions
from  Compton-thin  and  starburst  galaxies  is assumed  to  be  from
Compton-thick  AGN.  As a  result, our  model uses  more observational
information  to  constrain the  CT  AGN  fractions  as a  function  of
luminosity  and redshift,  including both  number counts  and redshift
distributions  at 24  $\mu$m.  In  our  model, we  allow the  redshift
evolution of the  CT AGN to be a free parameter,  while the CXB models
typically assume no evolution, fixed  evolution or an evolution as the
same as the Compton-thin AGN.

The  spectral  energy  distribution  (SED) of  individual  sources  is
crucial  to any  population model,  but our  joint model  and  the CXB
models  depend  on different  parts  of the  SED.   The  CXB model  is
sensitive  to  the  X-ray  SED above  10  keV  \citep[e.g.][]{Gilli07,
Treister09a}, while  ours relies on the 24$\mu$m/2-10keV  ratio of the
SMBH SED but  also the star-forming IR SED.  As a  result, we ran four
variants of the model  to incorporate the SED uncertainties, including
the reference  one, the  one with X-ray  to IR  ratio of the  SMBH SED
3-$\sigma$ (0.2 dex) above the average ratio, the one with X-ray to IR
ratio of the SMBH SED 3-$\sigma$ (0.2 dex) below the average ratio and
the  one assuming strong  redshift evolution  in the  star-forming SED
(for details, see Paper I).  Overall,  our model provides a new way to
constrain the  CT AGN  abundance using substantially  more information
from more  diverse deep survey  data.  Paper I presented  the detailed
model  construction and  three basic  outputs including  the  total IR
luminosity function,  the SMBH energy fraction  in the IR  band and HI
column  density distributions as  a function  of X-ray  luminosity and
redshift.

In this paper we discuss the predicted CT AGN abundance and compare it
to    a    large   collection    of    empirical   constraints.     In
\S~\ref{section_evl_CTAGN}, we  show the spatial number  density of CT
AGN.  We present the type-2 and CT AGN fraction as a function of X-ray
fluxes in  \S~\ref{section_agn_fraction}.  The contribution of
CT  AGN  to   the  SMBH  accretion  and  CXB   spectrum  is  shown  in
\S~\ref{section_SMBH_accretion}.   Discussions   and  conclusions  are
presented    in     \S~\ref{discussion}    and    \S~\ref{conclusion},
respectively.

\section{The Comoving Number Density Of Compton-Thick AGN}\label{section_evl_CTAGN}

\begin{figure}
\epsscale{1.2}
\plotone{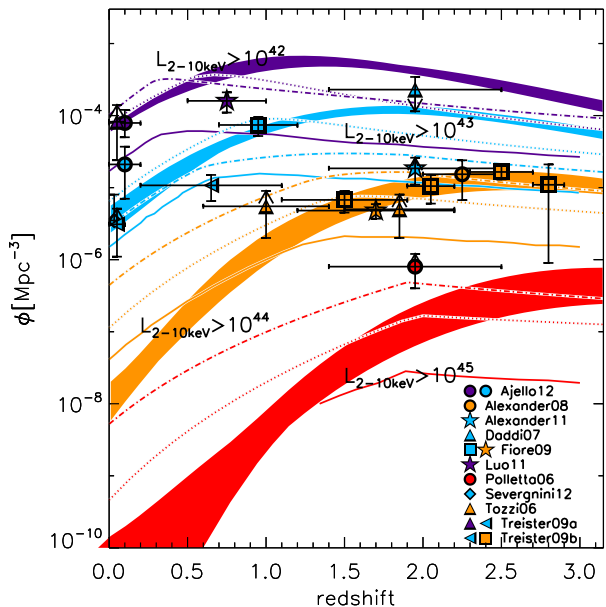}
\caption{\label{evl_CTAGN} Comoving number density of CT AGN above
different   intrinsic  2-10   keV  luminosities   as  a   function  of
redshift. The filled areas represent the predictions of our model in each of 
four luminosity bins. The width of the curves
reflect the uncertainties of the predictions. The solid, dotted and dotted-dashed
lines are  the predictions  by the cosmic  X-ray background  models of
\citet{Treister09a},     \citet{Gilli07}     and     \citet{Draper10},
respectively.  All  symbols represent  the observations.  A  symbol is
plotted  with a filled  core plus  a black  envelope.  For  all filled
areas,  lines  and  symbol  cores,  purple:  $L_{\rm  2-10  keV}$  $>$
10$^{42}$ erg/s, cyan: $L_{\rm 2-10 keV}$ $>$ 10$^{43}$ erg/s, orange:
$L_{\rm 2-10  keV}$ $>$ 10$^{44}$  erg/s, and red: $L_{\rm  2-10 keV}$
$>$ 10$^{45}$ erg/s.}
\end{figure}

 Figure~\ref{evl_CTAGN} shows the predicted comoving number density of
CT AGN above different intrinsic rest-frame 2-10 keV luminosity limits
as  a  function of  redshift  (filled  areas  with different  colors).
Symbols  with the  same color  are  the empirical  estimates from  the
literature  above the same  limits.  Among  different variants  of our
model, as  reflected by the vertical  width of the filled  area in the
figure, the predicted number density varies from a factor of 2 for low
luminosity objects ($>$ 10$^{42}$ erg/s)  up to a factor of 5
for  luminous ones ($>$  10$^{45}$ erg/s).   For CT  AGN with
intrinsic  $L_{\rm  rest  2-10keV}$  $>$ 10$^{42}$  erg/s,  our  model
predicts  the density to  peak at  a few  $\times$10$^{-4}$ Mpc$^{-3}$
around $z$ $\sim$ 1-1.5, declining slowly toward both higher and lower
redshifts.  The local densities as measured by \citet{Treister09a} and
\citet{Ajello12} are consistent with  our predictions.  The estimate at
$z$$\sim$0.7 by  \citet{Luo11} is  only a factor  of 2 lower  than our
prediction.  They  performed a Monte-Carlo simulation  to estimate the
CT AGN  fraction in  a sample of  IR-excess sources defined  as having
excess  IR emission  relative  to the  UV-based star-forming  emission
(log(SFR$_{\rm      IR+UV}$/SFR$_{\rm     UV,corr}$)      $>$     0.5)
\citep[e.g.][]{Daddi07}.  The stacked  X-ray spectrum of these sources
shows evidence for  heavy extinction.  However, due to  the lack of CT
signatures  in individual  galaxies, this  statistical  approach still
suffers from large uncertainties.

The more luminous CT AGN at $L_{\rm 2-10keV}$ $>$ 10$^{43}$ erg/s show
a different  trend, with  a faster evolution  starting at $z$=0  and a
higher peak redshift.  The  predicted local density is consistent with
empirical estimate by \citet{Treister09a} and \citet{Severgnini12} but
three  times  lower  than  that  by \citet{Ajello12}  yet  within  the
uncertainty.   The  work  by  \citet{Treister09a}  only  includes  the
transmission  AGN,  thus   underestimating  the  total  population  if
reflection-dominated CT  AGN are abundant. Beyond  the local universe,
the  density around $z$=0.7  by \citet{Treister09b}  is 3  times lower
than  the prediction  of  our  model, while  other  high-z studies  by
\citet{Fiore09}, \citet{Daddi07}  and \citet{Alexander11} give results
consistent with  the model. Among them, \citet{Daddi07}  gives a solid
upper-limit, as they assume CT  nature for all their IR-excess objects
(log(SFR$_{\rm   IR+UV}$/SFR$_{\rm    UV,corr}$)   $>$   0.5),   while
\citet{Alexander11}  derived  a  solid  lower-limit by  only  counting
sources that satisfy the BzK selection detected in the X-ray.

At $L_{\rm 2-10keV}$ $>$ 10$^{44}$  erg/s, the comoving density of the
CT AGN shows a rapid rise  until $z$$\sim$2 and almost a flat trend up
to  $z$=3.  The  model's prediction  is more  or less  consistent with
observations \citep{Alexander08, Fiore09,  Treister09b} except for one
data   point  around  $z$$\approx$0.9   but  within   the  uncertainty
\citep{Tozzi06}.   \citet{Tozzi06}  analysed   the  X-ray  spectra  of
sources in the  1Ms CDF-S and identified 14  CT AGNs, possibly missing
the  X-ray-undetected CT  AGNs.   At $L_{\rm  2-10keV}$ $>$  10$^{45}$
erg/s, the CT AGN show a rapid evolution in the model, with two orders
of  magnitude increase  in  density  from $z$=0  up  to $z$=3.   Model
prediction   around  $z$$\sim$2   is  lower   than  the   estimate  by
\citet{Polletta06}. Although only X-ray detected sources are accounted
for this measurement \citep{Treister09a}, the low number of objects caution a
large uncertainty associated with the measurement.

The predicted CT AGN number  density of our model is generally similar
to predictions  by the  CXB model of  \citet{Gilli07} (dotted  lines in
Figure~\ref{evl_CTAGN}) but ours peaks  at higher redshift.  The model
of  \citet{Gilli07} assumed all  X-ray sources  detected at  0.5-2 and
2-10  keV to  be Compton-thin  and  derived their  distributions as  a
function of  redshift, luminosity  and HI columns.   After subtracting
the contribution of these  Compton-thin sources from the CXB spectrum,
the residual is  not zero, thus implying the existence  of CT AGN.  By
assuming  the same  redshift  evolution  for CT  AGN  as for  obscured
Compton-thin AGN, the  amount of CT AGN is derived  by matching to the
CXB  spectrum.  At  a given  redshift and  limiting  X-ray luminosity,
their result does not  deviate from ours significantly.  A significant
difference however between the two models is that they predict a lower
redshift  for the  peak of  the CT  AGN number  density.   For $L_{\rm
2-10keV}$  $>$ 10$^{42}$  erg/s, our  peak redshift  is $z$$\sim$1-1.5
compared   to    theirs   around   $z\sim$0.7.     A   difference   of
$\Delta$$z$=0.5-1 in  the peak redshift  is also found for  the number
density of CT AGN at $L_{\rm 2-10keV}$ $>$ 10$^{43, 44, 45}$ erg/s.

The    CXB   model    of   \citet{Treister09a}    (solid    lines   in
Figure~\ref{evl_CTAGN})  predicts lower CT  AGN number  densities than
ours.   At $L_{\rm  2-10keV}$ $>$  10$^{42}$ erg/s,  our  predicted CT
density is  5-10 times  higher than theirs  across the  whole redshift
range.  At  higher luminosities ($>$  10$^{43, 44, 45}$  erg/s), their
predictions are  similar to  ours up to  their turn-over  redshift but
lower  by 5-10  at higher  redshifts as  our predicted  density rises
faster. This is not surprising, since Treister et al hold the CT  fraction in obscured AGN constant at
the local value seen in the INTEGRAL and Swift data, allowing for no redshift
evolution.   In contrast,  our model  requires more  CT AGN  at higher
redshifts to satisfy  the various sets of observations  (see Paper I).
As shown in the next section,  our model has no problem in reproducing
the CT AGN fraction as observed by INTEGRAL and Swift.

  In the  CXB model  of \citet{Draper10}  (dotted-dashed  lines in
Figure~\ref{evl_CTAGN}),  low luminosity CT  AGN ($>$  10$^{42}$ erg/s
and  $>$ 10$^{43}$ erg/s)  have comparable  number densities  to those
predicted from  our models at  low redshift, but are  noticeably lower
than ours at high $z$. The opposite  seems to be the case for the high
luminosity CT AGN,  where two models make similar  predictions at high
redshift, but differ at low redshift where the \citet{Draper10} models
predict more CT AGN than do our models.  These differences reflect the
weaker  dependence   of  their  models  on   luminosity  and  redshift
\citep{Draper10, Draper09}.

\section{The Type-2 and CT AGN Fraction}\label{section_agn_fraction}

\begin{figure*}
\epsscale{1.0}
\plotone{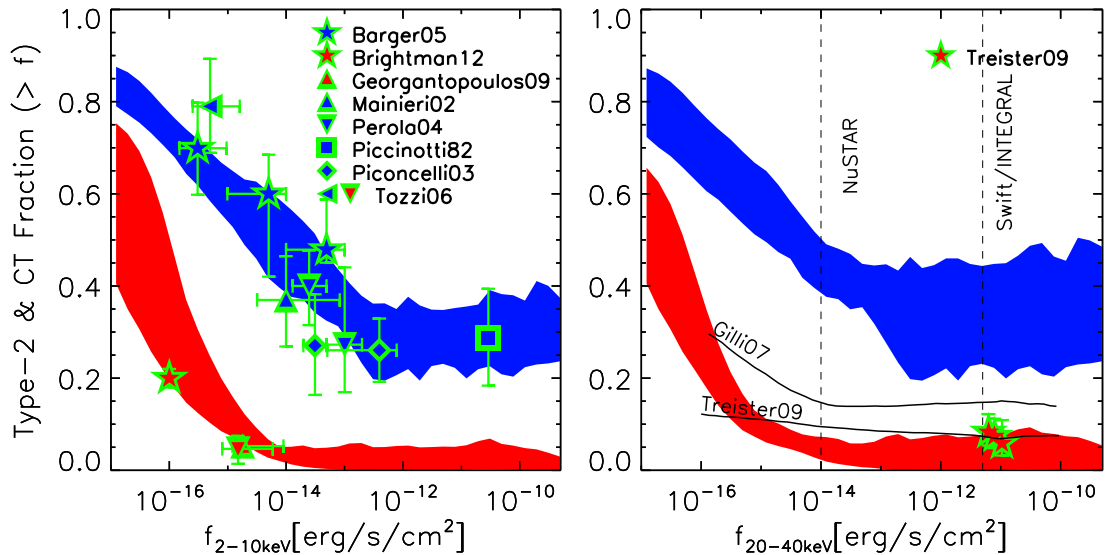}
\caption{\label{obsfrac_flux} Fraction  of type-2 (upper band) and
CT AGN (lower band) in all AGN  as a function of 2-10 kev (left panel)
and 20-40 keV (right panel)  fluxes.  The filled areas are our model's
predictions with  the vertical  width reflecting the  uncertainty. The
empirical estimates from the  literature are indicated by symbols that
are  plotted with  filled cores  plus green  outlines. For  all symbol
cores and  filled areas,  blue stands for  type-2 AGN  fraction (upper
band) while red is for the  CT AGN fraction (lower band). In the right
panel, two solid lines are the CT AGN fractions predicted by models of
\citet{Gilli07}  and  \citet{Treister09a}.    Dashed  lines  give  the
typical survey limits of NuSTAR, Swift and INTEGRAL. }
\end{figure*}

Figure~\ref{obsfrac_flux} plots  the fractions of AGN  that are type-2
($N_{\rm HI}$ $>$ 10$^{22}$  cm$^{-2}$) and Compton thick at different
2-10  keV  and  20-40  keV  fluxes. The  two  fractions  show  similar
behaviors, i.e., a flat trend at bright fluxes along with a rapid rise
toward  fainter ends.   Constant type-2  fractions of  30$\pm$10\% and
35$\pm$10\%  are found  at  2-10 keV  and  20-40 keV  above fluxes  of
10$^{-13}$  erg/s/cm$^{2}$, respectively, while  the CT  AGN fractions
remain around 3$\pm$3\% and 4$\pm$3\%  at 2-10 keV and 20-40 keV above
fluxes of  10$^{-15}$ erg/s/cm$^{-2}$,  respectively. In Paper  I, our
model predicts  a rapid  redshift evolution of  the type-2 and  CT AGN
fraction  at  given  intrinsic  X-ray  luminosities,  making  the  two
fractions increase with decreasing fluxes, which combined with further
obscuration to  CT/type-2 objects  results in a  flat trend  at bright
fluxes but a rapid rise toward lower fluxes.

As  shown  in the  figure,  the  predicted  type-2 AGN  fractions  are
consistent  with empirical  constraints  as compiled  in  the work  of
\citet{Gilli07}    including   \citet{Barger05},   \citet{Mainieri02},
\citet{Perola04},   \citet{Piccinotti82},   \citet{Piconcelli03},  and
\citet{Tozzi06}.   All these  studies identified  the type-2  in X-ray
flux limited samples  through X-ray spectral analysis. For  the CT AGN
fraction  as  a  function  of  the  2-10 keV  flux,  we  compared  our
predictions  to empirical estimates  based on  studies of  three X-ray
flux  limited samples \citep{Tozzi06,  Georgantopoulos09, Brightman12}
that  are constructed  from Chandra  1 Ms,  2Ms and  4Ms  survey data,
respectively.  The first two  identified CT AGN through X-ray spectral
analysis  and  derived  CT  fractions  of 5\%  down  to  $f_{\rm  2-10
keV}$=10$^{-15}$  erg/s/cm$^{2}$, consistent  with the  predictions of
our model. The majority of  these objects are CT AGN whose transmitted
light   dominates    over   the   reflected    radiation   in   X-ray.
\cite{Brightman12} carried  out X-ray  spectral analysis of  449 X-ray
sources down to  a flux 10 times lower  ($f_{\rm 2-10 keV}$=10$^{-16}$
erg/s/cm$^{2}$), with average photon counts 3-5 times smaller than the
other  two studies.   After corrections  for these  two  effects, they
argued  for  20$\pm$2\%  CT  AGN  fraction,  lower  than  our  model's
prediction but  within the uncertainty.   At 20-40 keV, our  result is
consistent with  the CT fraction as identified  in Swift/INTEGRAL data
\citep{Treister09a} in which all CT AGN are identified as transmission
sources.   The recently  launched NuSTAR  mission should  offer strong
constraints on the CT AGN fraction down to 10$^{-14}$ erg/s/cm$^{2}$.

As already discussed in the  previous section, our model predicts roughly
the  same CT AGN  abundance as  the model  of \citet{Gilli07}  but our
predicted CT AGN number density peaks at higher redshift. This is also
reflected  in  Figure~\ref{obsfrac_flux} (right  panel)  where our  CT
fraction  is lower  than  theirs above  10$^{-15}$ erg/s/cm$^{2}$  but
exceeds  theirs at  fainter flux  levels.   Compared to  the model  of
\citet{Treister09a}, our model predicts more abundant CT AGN 
at high $z$, resulting in a similar CT fraction
above 10$^{-15}$ erg/s/cm$^{2}$ but a significantly higher fraction in
our model at fainter fluxes.

\section{The Contribution Of CT AGN To SMBH Growth And CXB}\label{section_SMBH_accretion}

\begin{figure}
\epsscale{1.1}
\plotone{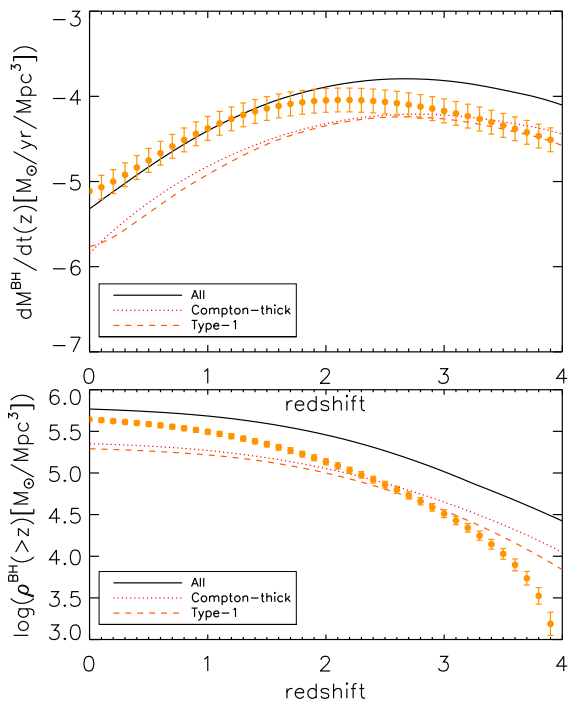}
\caption{\label{SMBH_accretion}  Comoving  SMBH  accretion  rate
(upper panel) and total accreted  SMBH mass density (lower panel)
for  all AGN (solid  lines), type-1  (dashed lines)  and Compton-thick
(dotted lines) predicted from our model. Each line is the median value
of  the predictions from  four model variants (see Paper I), while the  
uncertainty of the predictions among model variants is around
50\%. Symbols are from \citet{Hopkins07}. Our predicted  accretion rates  seem to
match  those presented  in  \citet{Hopkins07} up  to  $z$=2, but  exceed
theirs at higher redshifts, while the discrepancy in the cumulative accreted SMBH  mass across the
redshift is caused by that above $z$=2.}
\end{figure}

An SMBH  grows through accretion, and the accretion disk is responsible for the optical through X-ray emission.   
By introducing  the  radiative  efficiency
$\epsilon_{r}$, the  observed SMBH  luminosity and accretion  rate are
related by:
\begin{equation}
 L_{\rm ox}^{'} = \epsilon_{r}\dot{M}_{\rm BH}c^{2}
\end{equation}  
where $L_{\rm  ox}^{'}$ is the luminosity integrated  from the optical
to the  X-ray (1  $\mu$m to  200 keV), $\epsilon_{r}$  is the  mass to
energy conversion rate, $\dot{M}_{\rm BH}$ is SMBH mass accretion rate
and $c$ is the speed of  light.  As constructed in Paper I, our quasar
SED invokes the luminosity-dependent  optical to X-ray ratio, and thus
the correction  from 2-10 keV to  the luminosity from 1  $\mu$m to 200
keV is luminosity  dependent with an average value  around a factor of
30.   Figure~\ref{SMBH_accretion} shows  the  comoving SMBH  accretion
rate (upper  panel) and  cumulative accreted SMBH  mass above  a given
redshift  (lower panel). Our predicted  accretion rates  seem to
match  those presented  in  \citet{Hopkins07} up  to  $z$=2, but  exceed
theirs at higher redshifts, while the discrepancy in the cumulative accreted SMBH  mass across the
redshift is caused by that above $z$=2.  We  believe the discrepancy to be related
to the large  uncertainties in deriving the obscured  AGN LFs at these
redshifts.      The   derived   local   SMBH    mass   density   is
(5.8$\pm$1.0)$\times$10$^{5}$     M$_{\odot}$     Mpc$^{-3}$     given
$\epsilon_{r}$=0.1. This number is quite consistent with those derived
from   local   bulge   mass   functions  through   the   bulge/BH-mass
relationship,   (2.9$\pm$0.5)$\times$10$^{5}$  M$_{\odot}$  Mpc$^{-3}$
\citep{Yu02},   (4.2$\pm$1.1)$\times$10$^{5}$  M$_{\odot}$  Mpc$^{-3}$
\citep{Shankar04}  and $4.6^{+1.9}_{-1.4}$$\times$10$^{5}$ M$_{\odot}$
Mpc$^{-3}$  \citep{Marconi04}.  Our model  further predicts  that only
33\% of local  SMBH mass is accreted in the type-1  phase while the CT
accretion contributes as much as 38\% to the local SMBH mass density.

\begin{figure}
\epsscale{1.1}
\plotone{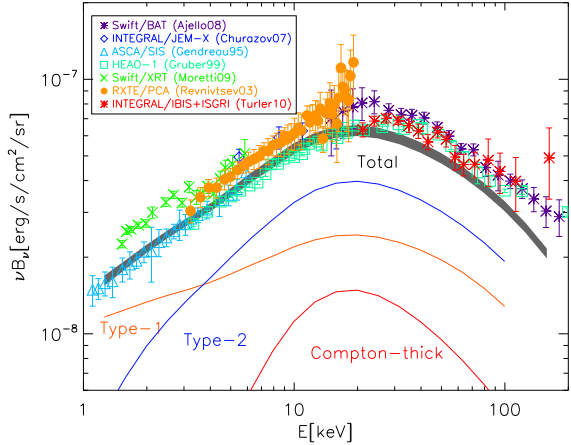}
\caption{   \label{cosmic_xray_light}  The  cosmic   X-ray  background
spectrum from our model compared to    the   observations  from   the
literature (all symbols).  The grey curve is the prediction from our model, with the thickness
of the curve representing the uncertainty in the model as a function of energy. 
Colored lines represent  the median  values  of predictions  by four  model
variants (see Paper I) for type-1 (orange), type-2 (blue) and Compton-thick (red) AGN as a
function of energy.
References:    Ajello08    --    \citet{Ajello08};    Churazov07    --
\citet{Churazov07};  Gendreau95  --  \citet{Gendreau95};  Gruber99  --
\citet{Gruber99};  Moretti09  --  \citet{Moretti09};  Revnivtsev03  --
\citet{Revnivtsev03}; Turler10 -- \citet{Turler10}. }
\end{figure}

Figure~\ref{cosmic_xray_light}  shows  our   prediction  for  the  CXB
spectrum at 1-200 keV as compared to observations.  Our model predicts
a  peak around  20  keV.  Below  20  keV, our  prediction matches  the
results     of    RXTE/PCA    \citep{Revnivtsev03}     and    ASCA/SIS
\citep{Gendreau95}, but  is about 20\%  lower than those  of Swift/XRT
\citep{Moretti09}  and  INTEGRAL/JEM-X  \citep{Churazov07}.    At
20-100  keV, the  model is  systematically lower  by 20-30\%  than the
observations  \citep{Gruber99, Ajello08,  Turler10}.  We  do  not know
exactly  what causes the discrepancy  but noticed  that  the model  cannot
re-produce the local 15-55 keV  counts of \citet{Ajello12} and the CXB
spectrum  above 20  keV  at the  same  time.  If  the  model fits  the
\citet{Ajello12} counts,  it under-produces the CXB  spectrum above 20
keV.  On the  other  hand,  if the  model  is forced  to  fit the  CXB
spectrum, it would over-predict  counts of \citet{Ajello12}.  As noted
in  \citet{Ajello12},  previous CXB  models  have  a similar  problem,
including  \citet{Gilli07},  \citet{Treister09a} and  \cite{Draper10},
where they fit the CXB  spectrum well but over-predict local 15-55 keV
counts.  A key  observational input  for the  CXB models  to  fit both
\citet{Ajello08}  counts  and  CXB   spectrum  above  20  keV  is  the
rest-frame  SED at  energies  above 20  keV  which is  still not  well
constrained given limited observations. But certainly we cannot  exclude the possibility 
that our model has limitations in reproducing the CXB spectrum above 20 keV as it fits so many data points
over a very large range of the frequency (X-ray and IR/submm) to minimize $\chi^{2}$.

As shown in the figure, type-1 AGN dominate the CXB
below 5 keV,  above which the type-2 are  responsible for the majority
of  CXB emission.   The CT  AGN  contribution rises  quickly from  low
energy and peaks around $\sim$25\% at 20 keV.

\section{Discussion}\label{discussion}

\begin{figure}
\epsscale{1.1}
\plotone{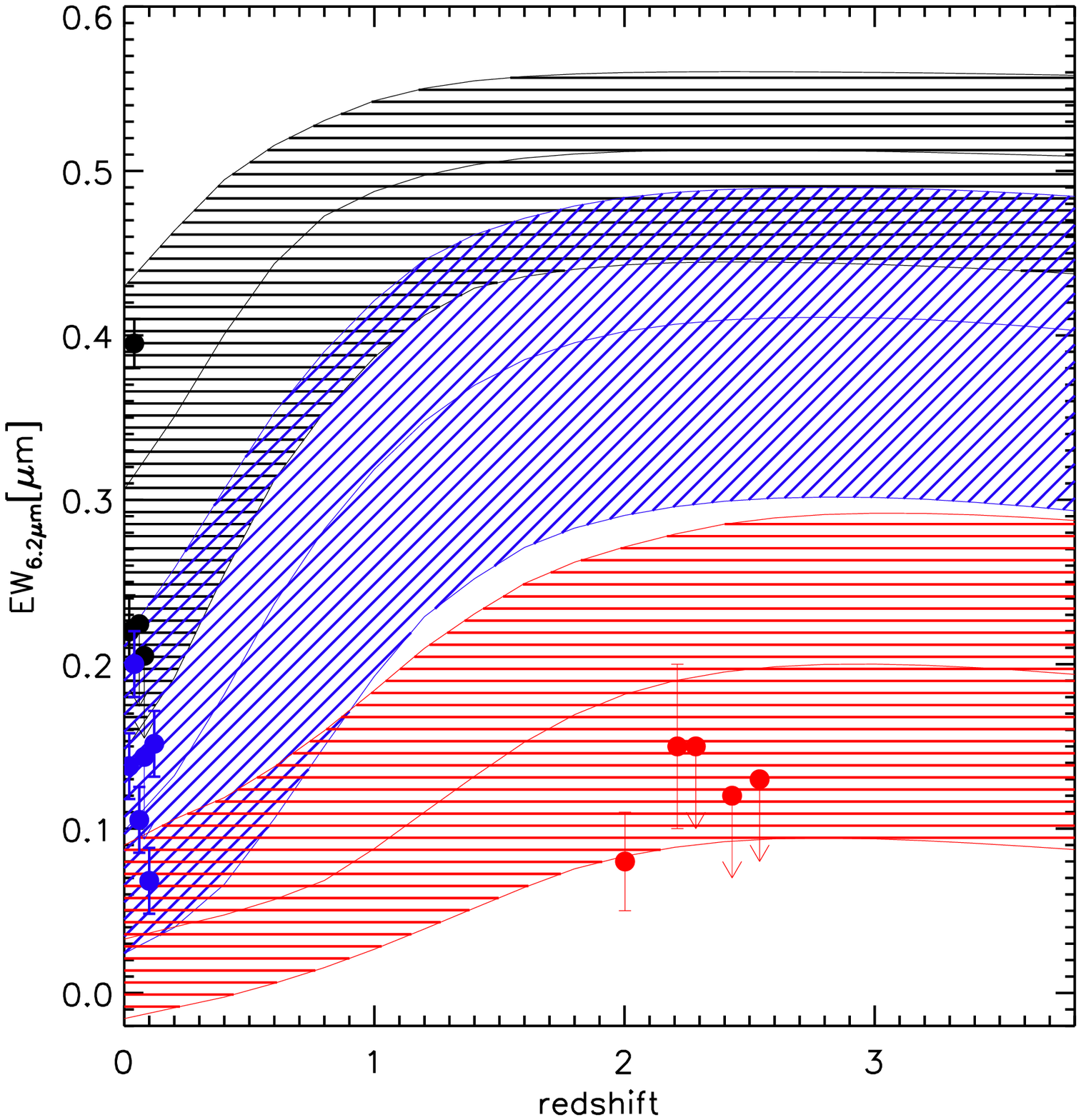}
\caption{\label{ew_CT_lx_red}  Predicted  median EW$_{\rm 6.2{\mu}mPAH}$
 and 20-80\% probability  range of CT AGN in three intrinsic
X-ray luminosity  ranges.  Symbols  are observations of low-z  \citep{Risaliti99} 
and high-z 
\citep{Alexander08} CT AGN.   For  all   shaded  area   and   symbols,  black:
$L_{\rm 2-10keV}$=10$^{42}$-10$^{43}$ erg/s, blue: $L_{\rm 2-10keV}$=
10$^{43}$-10$^{44}$ erg/s and red: $L_{\rm 2-10keV}$ $>$ 10$^{44}$
erg/s. Note  that in our  model, pure star  formation has a EW  of 0.6
$\mu$m;  any  value  below  that  is due  to  contributions  from  AGN
emission. }
\end{figure}

Our  model  predicts  an  abundant  CT AGN  population  especially  at
high-$z$.    By  comparing   Figure~\ref{evl_CTAGN}   to  the   type-1
unobscured AGN  number density as measured  by \citet{Hasinger05}, our
predicted CT AGN density is 3-5 times higher at $L_{\rm 2-10 keV}$ $>$
10$^{42}$  erg/s.  At  $L_{\rm  2-10 keV}$  $>$  10$^{44}$ erg/s,  the
predicted  CT AGN  density is  still  comparable to  their type-1  AGN
around  $z\sim$  2.   As shown  in  Paper  I,  our model  predicts  an
increasing CT AGN fraction with redshift, resulting in a larger CT AGN
population   at   high-$z$  as   compared   to   the  predictions   of
\citet{Gilli07}  and \citet{Treister09a}.  The large  CT  AGN fraction
around $z$ $\sim$ 2 may  be consistent with observational evidence for
high gas fractions  and associated high SFRs of  $z$$ \sim$ 2 galaxies
\citep{Tacconi10}.  The high velocity dispersion  of $z$ $ \sim$ 2 gas
disks  implies  a  large  vertical  height  \citep{ForsterSchreiber06,
Law07, Genzel08}, as might result from continuous stirring by on-going
star formation  \citep{Elmegreen10}. If such behavior  persists as gas
is transported down to the central 1-10 pc scale around the nuclear BH
\citep{Wada02, Hopkins12}, the dusty  torus of high-$z$ AGN might have
a larger vertical extent and  subsequently cover a larger solid angle,
resulting in a larger CT  AGN fraction at high-z \citep{Fabian99}.  In
spite   of    their   abundance   in   our   model,    as   shown   in
Figure~\ref{obsfrac_flux}, CT  AGN only dominate  at $f_{\rm 2-10keV}$
$<$  10$^{-15}$ erg/s/cm$^{2}$,  where little  data is  available from
current X-ray  missions to  reliably identify these  objects.  Another
possibility to explain the  increasing obscured fraction is related to
redshift evolution of major  mergers as shown by \citet{Treister10} in
which abundant dust and gas  brought in by mergers obscure the nucleus
before they  are disrupted by  radiation pressure to reveal  a type-1,
unobscured quasar phase.


\begin{figure}
\epsscale{1.2}
\plotone{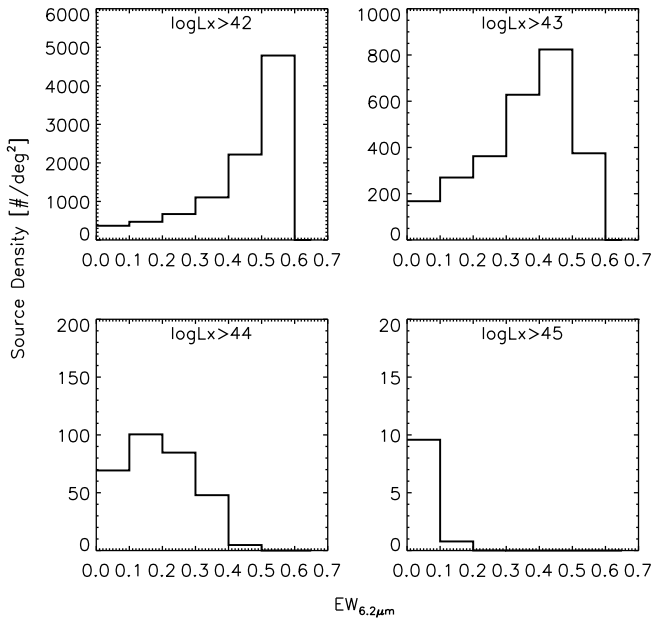}
\caption{\label{ew_dist_CT}  Predicted  distribution  of  6.2
$\mu$m  aromatic  feature EW  for  CT  AGN  above different  intrinsic
rest-frame 2-10 keV luminosities.  In our model, pure star
formation has a EW of 0.6 $\mu$m; any value below that is due to contributions 
from AGN emission.}
\end{figure}

\begin{figure}
\epsscale{1.2}
\plotone{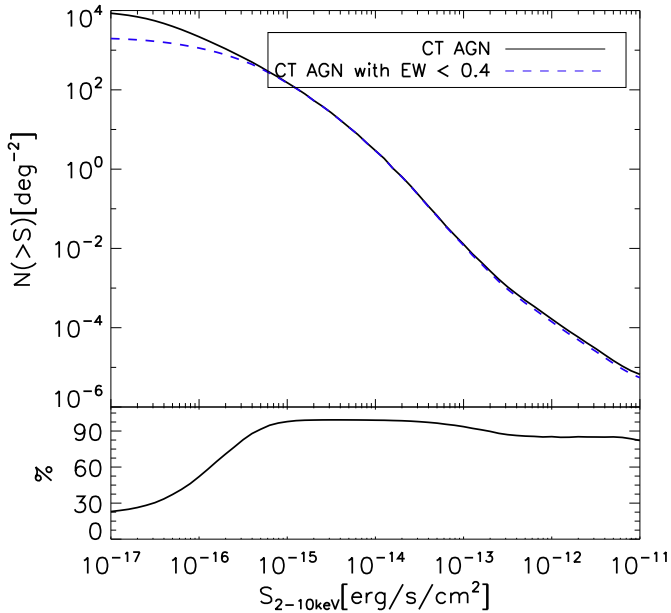}
\caption{\label{lowEW_CT_fx} Cumulative surface density of CT AGN (intrinsic rest-frame 
2-10 keV luminosity above 10$^{42}$ erg/s and HI column density above 10$^{24}$ cm$^{-2}$) and CT AGN
with 6.2 $\mu$m aromatic feature EW lower than 0.4 (from which the AGN mid-IR featureless emission can be detected).}
\end{figure}

The featureless mid-IR emission from the dusty torus of AGN has been a
powerful  way  to  infer  the  intrinsic  accretion  luminosity.  
Especially when appearing  along with weak or no  X-ray emission, this
continuum offers  a strong indicator  of CT HI  columns \citep{Lacy04,
Stern05,       Alonso-Herrero06,      Treister09b,      Alexander08}.
Figure~\ref{ew_CT_lx_red}  gives  the  predicted  median  and  20-80\%
probability range of 6.2 $\mu$m aromatic feature EW of CT AGN in three
intrinsic   X-ray   luminosity   ranges,  10$^{42}$-10$^{43}$   erg/s,
10$^{43}$-10$^{44}$ erg/s  and $>$ 10$^{44}$ erg/s.  Note  that the EW
of star-forming templates  in our model has a value  of 0.6 $\mu$m, so
smaller  values indicate  the presence  of the  contribution  from the
dusty torus.  In Paper  I, we compared our model's predictions on
the EW  distributions to observations for two  Spitzer legacy programs
(GOALS    \&    5MUSES)    and    found   a    general    consistency.
Figure~\ref{ew_CT_lx_red} also gives the  observed EW of individual CT
AGN drawn from the literature,  where  the local  sample  is  from
\citet{Risaliti99},      and      high-z      sample      is      from
\citet{Alexander08}. Although a fair  comparison between the model and
observation is impossible  for such a sample, the  data are within the
model predicted range.   Figure~\ref{ew_dist_CT} further shows the 6.2
$\mu$m feature EW distribution  of AGN above different intrinsic X-ray
luminosities.  We define objects with EW$_{6.2 {\mu}m}$ $<$ 0.4 $\mu$m
as  those  with  detectable   hot  dust  emission  from  AGN,  because
observations confirm  a small scatter  in the star-forming  6.2 $\mu$m
EWs, with  the median of  0.6 $\mu$m and 3-$\sigma$  dispersions about
0.2-0.25   $\mu$m  \citep[e.g.][]{Wu10,   Stierwalt13}.    The  figure
indicates that the fractions of CT AGN that have EW$_{6.2 {\mu}m}$ $<$
0.4, i.e.,  detectable AGN mid-IR  emission, are 27\%, 55\%,  98\% and
100\% for  the limiting intrinsic rest-frame 2-10  keV luminosities of
10$^{42}$,  10$^{43}$, 10$^{44}$  and  10$^{45}$ erg/s,  respectively.
Thus  almost  all  intrinsically  luminous CT  AGN  (unobscured  $L_{\rm
rest-2-10keV}$  $>$  10$^{44}$   erg/s)  have  detectable  AGN  mid-IR
emission,  consistent with recent  IR studies  of high  luminosity AGN
\citep{Mateos13}.  Figure~\ref{lowEW_CT_fx}  shows the fraction  of CT
AGN with EW$_{6.2{\mu}m}$ $<$ 0.4 $\mu$m as a function of the 2-10 keV
flux.  Above  10$^{-15}$ erg/s/cm$^{2}$ where intrinsic  bright CT AGN
dominate,  80-90\% of  CT AGN  have EW$_{6.2{\mu}m}$  $<$  0.4 $\mu$m.
Below  10$^{-15}$ erg/s/cm$^{2}$, the  fraction drops  with decreasing
fluxes but still remains appreciable, around 60\% and 30\% at 2-10 keV
flux  of  10$^{-16}$  erg/s/cm$^{2}$  and  10$^{-17}$  erg/s/cm$^{2}$,
respectively.

\begin{figure}
\epsscale{1.2}
\plotone{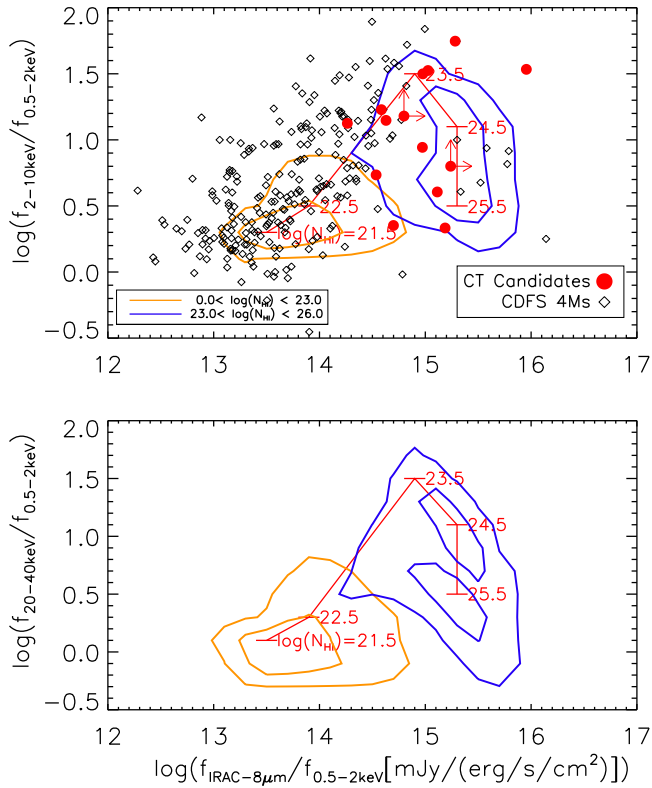}
\caption{\label{ir_xray_hx_sx} Distribution of  AGN with different HI
columns in  the X-ray hardness  vs. IR/X-ray ratio  plane: log$(f_{\rm
2-10keV}/f_{\rm   0.5-2keV})$  vs.   log$(f_{\rm  IRAC-8{\mu}m}/f_{\rm
0.5-2keV})$  for  the  upper  panel and  log$(f_{\rm  20-40keV}/f_{\rm
0.5-2keV})$ vs.  log$(f_{\rm IRAC-8{\mu}m}/f_{\rm 0.5-2keV})$  for the
lower  panel.   Orange  and  green  contours are  for  unobscured  and
mildly-obscured AGN (log$N_{\rm HI}$  $<$ 23) and heavily-obscured AGN
(log$N_{\rm HI}$  $>$ 23), respectively. For each  contour, two levels
enclose 60\%  and 90\% of  objects, respectively. The  diamond symbols
give  distributions of  all CDF-S  sources \citep{Xue11}  while filled
circles are CT AGN candidates from \citet{Alexander08, Alexander11}.}
\end{figure}

 The reliable  identification of CT  AGN beyond the local  universe is
still a challenge, even though multi-wavelength tools that combine the
X-ray data  with optical/IR photometry  have been developed  to reveal
many candidates.  The  upper panel of Figure~\ref{ir_xray_hx_sx} shows
the distribution  of AGN in  the plane of  log($f_{\rm 2-10keV}/f_{\rm
0.5-2keV}$)  vs.   log($f_{\rm  IRAC8{\mu}m}/f_{\rm 0.5-2keV}$).   The
median positions  of AGN with  log($N_{\rm HI}$/cm$^{-2}$)=21.5, 22.5,
23.5, 24.5 and  25.5 are labelled. Heavily obscured  AGN ($N_{\rm HI}$
$>$ 10$^{23}$ cm$^{-2}$)  are predicted to be well  separated from the
unobscured  and mildly  obscured objects  ($N_{\rm HI}$  $<$ 10$^{23}$
cm$^{-2}$).  Note that in our  model, the range in the distribution of
a  given   $N_{\rm  HI}$  is   mainly  caused  by  variation   in  the
redshift-dependent K correction and contamination by star formation in
the IRAC-8$\mu$m band. It does not incorporate the effect of variation
in  the  AGN  SED  at   X-ray  and  IR  wavelengths.   Also  shown  in
Figure~\ref{ir_xray_hx_sx} are the locations  of the CDFS sources from
\citet{Xue11},  which  spread  over  a  large region  in  this  plane.
However, the  distributions of observed CT AGN  candidates as compiled
from \citet{Alexander08, Alexander11} roughly lie within our predicted
range   of   heavily-obscured    AGN,   supporting   the   idea   that
heavily-obscured AGN can be identified through this simple diagnostic.
The validity  of a similar diagnostic  plot has also  been proposed by
\citet{Severgnini12}   based   on   photometry  of   local   confirmed
Compton-thick     AGN.     In    the     lower     panel    of
Figure~\ref{ir_xray_hx_sx},   we   also   examine  the   validity   of
log($f_{\rm    20-40keV}/f_{\rm     0.5-2keV}$)    vs.     log($f_{\rm
IRAC8{\mu}m}/f_{\rm  0.5-2keV}$) in separating  AGN with  different HI
columns, which has a similar  ability to the above diagnostic as shown
in the upper panel.

\begin{figure}
\epsscale{1.1}
\plotone{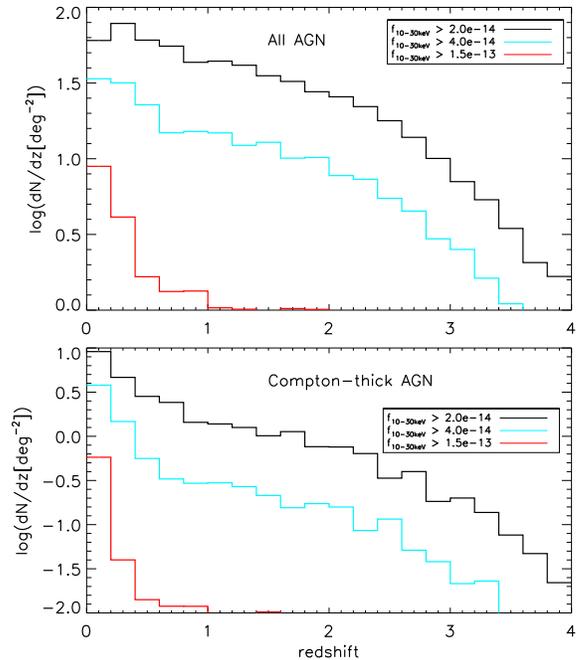}
\caption{\label{NuSTAR}  Predicted redshift distribution of NuSTAR AGN (upper panel) and NuSTAR CT AGN (lower panel)
for three survey depths as listed in \citet{Harrison13}.}
\end{figure}

Launched in  June 2012, NuSTAR \citep{Harrison13}  should quickly help
constrain the  distribution of CT AGN.   Figure~\ref{NuSTAR} shows the
predicted redshift distribution of all AGN and CT AGN above three flux
limits   targeted  in  NuSTAR   surveys,  namely   $f_{10-30keV}$  $>$
2$\times$10$^{-14}$,   4$\times$10$^{-14}$  and  1.5$\times$10$^{-13}$
erg/s/cm$^{2}$.  Given  0.3, 1-2 and  3 square degree for  those three
flux limits,  respectively, our model predicts $\sim$100  AGN in total
but only a  few CT AGN that  will be detected by NuSTAR  at 10-30 keV.
Our predicted total number of  AGN is similar to to recent predictions
by \citet{Ballantyne11}, but  our predicted number of CT  AGN is lower
than  theirs with  the  amount of  the  difference (a  factor of  2-5)
depending on the X-ray LFs they adopted.

\section{Conclusions}\label{conclusion}

We  employ our  joint population  model of  X-ray and  IR  backgrounds to
predict  the CT  AGN abundance  and  compare it  to a  diverse set  of
empirical determinations.  The main conclusions are:

(1) At intrinsic $L_{\rm rest 2-10 keV}$ $>$ 10$^{42}$ erg/s,
the CT AGN  density is predicted to be  around a few $\times$10$^{-4}$
Mpc$^{-3}$.  The density of higher luminosity CT AGN increases rapidly
from $z$=0 to $z$$\sim$2-3 and peaks at higher $z$.

(2)  The  CT AGN  fraction  appears  to be  low  (2-5\%  ) at  $f_{\rm
2-10keV}$  $>$  10$^{-15}$  erg/s/cm$^{2}$  but increases  rapidly  at
fainter flux levels.

(3)  The SMBH  accretion in  CT  AGN accounts  for 38\%  of the  total
accreted SMBH mass and contributes to  25\% of the CXB spectrum at its
peak.

(4)  We  also  investigate the  mid-IR  spectra  of  CT AGN  based  on
techniques that have been developed  to identify CT objects. The model
predicts that the majority (90\%)  of bright CT AGN ($L_{\rm rest 2-10
keV}$ $>$ 10$^{44}$ erg/s or $f_{\rm 2-10keV}$ $>$ 10$^{-15}$
erg/s/cm$^{2}$) have detectable hot dust emission from dusty tori; the
fraction drops for  faint objects, reaching 30\% at  $L_{\rm rest 2-10
keV}$ $>$ 10$^{42}$ erg/s or $f_{\rm 2-10keV}$ $>$ 10$^{-17}$
erg/s/cm$^{2}$.  Based  on  this,  we    confirm  that heavily obscured AGN
($N_{\rm HI}$ $>$ 10$^{23}$ cm$^{-2}$) can be separated from lower HI column AGN
through the plane of  the observed-frame X-ray hardness vs. mid-IR/xray ratio.

\section{Acknowledgment}

We  thank the  anonymous  referee for  helpful comments that improve 
the paper significantly.     The work is  supported through  the Spitzer  5MUSES Legacy
Program 40539. The authors  acknowledge support by NASA through awards
issued by JPL/Caltech.

\end{document}